# Robust Nonlinear Functional Link Minimum Total Error Entropy Adaptive Filtering Algorithm for Nonlinear Active Noise Control

Yuan Gao[a], Haiquan Zhao[a,*]

*a: Key Laboratory of Magnetic Suspension Technology and Maglev Vehicle, Southwest Jiaotong University, Chengdu, China.*

**Abstract**

Considering the distortions introduced in signal transmission channels under the Errors-in-Variables (EIV) model, the Total Least Squares (TLS) method has been widely adopted due to its ability to minimize the squared total error. However, TLS tends to underperform in non-Gaussian noise environments, as it fails to fully exploit the statistical characteristics of the noise. To address this limitation, a novel Minimum Total Error Entropy (MTEE) criterion is proposed by combining the strengths of the TLS and Minimum Error Entropy (MEE) frameworks, enabling more effective handling of the EIV model under non-Gaussian noise conditions. Building on the MTEE criterion, a Functional Link Artificial Neural Network (FLANN) structure is employed to model the nonlinear characteristics of the system, resulting in the development of the MTEE-FLANN algorithm. Furthermore, recognizing the high computational complexity typically associated with MEE-type algorithms, a quantized variant, referred to as the QMTEE-FLANN algorithm, is proposed to significantly reduce computational demands while maintaining performance. In addition, theoretical analyses regarding the convergence behavior, step-size stability range, and computational complexity of the proposed algorithms are presented to support their practical deployment. To validate the applicability and robustness of the developed methods, experiments are conducted under a nonlinear Active Noise Control (ANC) framework based on the EIV model. The results demonstrate the superior performance and strong robustness of the proposed algorithms compared to existing approaches.

## 1. INTRODUCTION

With the rapid advancement of modern society and technology, noise pollution has emerged as a significant global concern, posing substantial threats to public health and quality of life [1]–[6], [27]-[32], [34]. Prolonged exposure to noisy environments can lead to various health problems, including tinnitus, hearing loss, cardiovascular diseases, and, in severe cases, mortality. Therefore, effective noise control and mitigation strategies are urgently needed.

Noise control approaches are traditionally categorized into three main strategies: source control, propagation path management, and protection of individuals. Passive Noise Control (PNC) techniques [2], relying on acoustic materials to absorb sound energy [3], [4], are effective for mid-to-high frequency noise but become impractical for low-frequency noise due to their reliance on bulky, heavy, and expensive structures. To address these limitations, Active Noise Control (ANC) systems have been developed [5], [6], utilizing adaptive signal processing techniques to generate anti-noise signals that destructively interfere with primary noise.

However, conventional ANC systems predominantly assume linear models, whereas real-world environments often exhibit significant nonlinearities arising from actuator dynamics, acoustic propagation, and hardware imperfections [7]–[11]. Additionally, traditional ANC frameworks typically presume noise-free input signals, an assumption often violated in practice due to sensor inaccuracies, environmental disturbances, and sampling errors [12]. These factors lead to nonlinear distortions, biased estimations, and degraded system performance, highlighting the need for more robust and flexible ANC modeling approaches.

The Total Least Squares (TLS) criterion [13] has been widely used to account for errors in both input and output signals under the Errors-in-Variables (EIV) model. However, TLS inherently assumes Gaussian noise and focuses on minimizing the Euclidean norm of total errors, which limits its effectiveness in non-Gaussian environments. In contrast, the Minimum Error Entropy (MEE) criterion [14] provides superior robustness against non-Gaussian noise by minimizing information-theoretic measures of error distributions. Nonetheless, conventional

MEE-based algorithms do not explicitly address the input noise issue, often leading to suboptimal performance under the EIV framework.

To simultaneously address input-output noise contamination and non-Gaussianity in nonlinear systems, this paper proposes a novel Minimum Total Error Entropy (MTEE) criterion. Unlike traditional MEE or TLS approaches, MTEE inherently considers input-output noise jointly under a unified information-theoretic framework, making it particularly suitable for EIV scenarios where both the input and output measurements are corrupted.

Moreover, although deep neural networks and kernel-based filtering methods have demonstrated strong nonlinear modeling capabilities, they are often associated with high computational complexity, large parameter sets, and the need for extensive training data. These characteristics make them less suitable for real-time ANC applications, where low latency and computational efficiency are critical. In contrast, Functionally Linked Artificial Neural Networks (FLANNs) offer a favorable balance between modeling capacity and simplicity [8]-[11], [33], [35],[36]. By utilizing explicit basis function expansions, such as triangular basis functions, FLANNs can efficiently capture nonlinear mappings with significantly reduced computational overhead and without the risk of overfitting.

Given the practical considerations, the EIV-based ANC model is adopted as the experimental framework for algorithm validation. This choice reflects the reality that in practical ANC systems, both input signals (reference microphone signals) and output signals (error microphone signals) are contaminated by various noise sources. The EIV framework provides a more realistic and challenging testbed compared to conventional ANC models, ensuring that the proposed algorithms are robust and applicable in practical deployments.

The main contributions of this paper are outlined below:

1)Nonlinear Distortion in the EIV Model: To handle the nonlinear distortion in the signal transmission channel under the EIV model and non-Gaussian noise interference, we introduce the MEE and TLS criteria to handle the non-Gaussian input-output noise. A FLANN structure with triangular basis expansion is adopted to effectively capture system nonlinearity, leading to the development of the MTEE-FLANN algorithm.

2)Algorithm improvement: To handle the computational complexity associated with the MEE-type algorithm, an improved quantized method is utilized and the QMTEE-FLANN algorithm is proposed. This method reduces computational burden significantly by considering only the relationship between each instantaneous error and the local mean error within a sliding window, making the algorithm nearly independent of the error sample size.

3)Performance analysis: The local extreme point, local stability step stability range, and computational complexity of these noval algorithms are further introduced, which providing a strong theoretical foundation for practical implementations.

4) ANC model in EIV and algorithm validation: Traditional ANC system is designed for linear models, and does not take into account the input and output noise content. Consider the more common nonlinear EIV models in real scenarios, these noval algorithms are applied to nonlinear ANC. The corresponding FsFLMTEE and FsFLQMTEE algorithms are proposed. The results indicate the robustness of the presented algorithms under the EIV model.

The content of this article is arranged below: Section 2 introduces the FLANN model and TLS criterion. In Section 3, the MTEE-FLANN is derived in detail, and the QMTEE-FLANN is proposed to address the computational complexity. Section 4 and 5 analyzes the convergence performance and computational complexity of these algorithms under certain simplifying assumptions. Section 6 offers numerical simulations and evaluates the performance of these presented algorithms in the context of a novel nonlinear ANC model. Finally, Section 7 concludes this paper.

## 2. Background

### *2.1. Functional Link Artificial Neural Network*

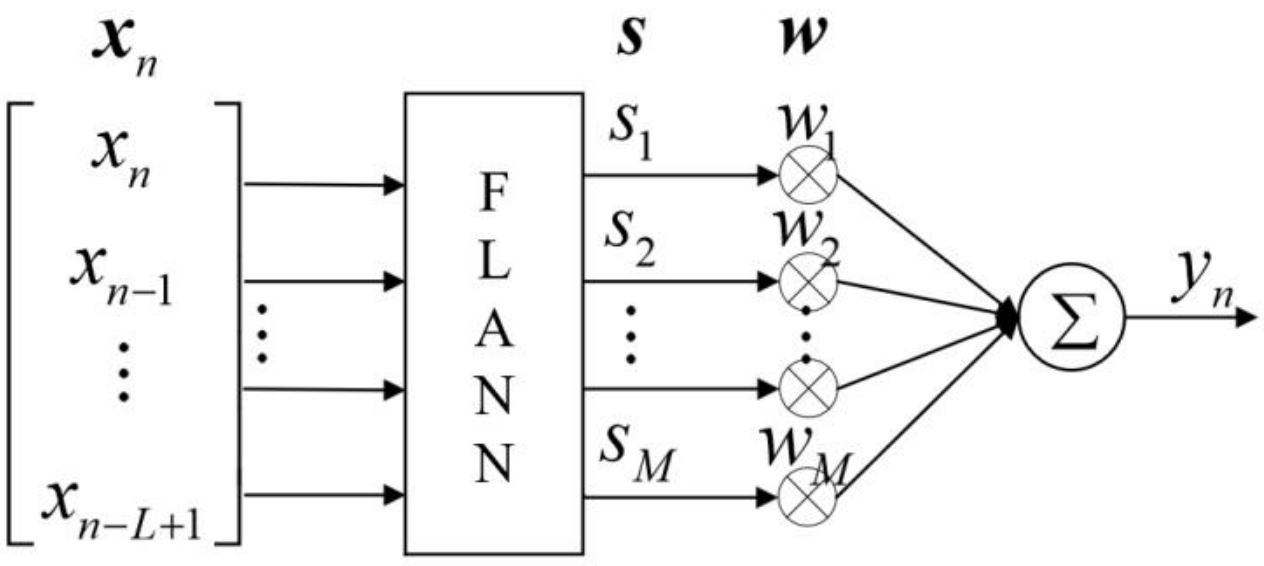

/

Figure. 1. The structure of the FLANNs

FLANN uses a single-layer learning network that does not contain hidden layers. In Fig. 1, the L-item input signal is expanded into M-items by a function, and the expanded signal is then linearly combined with adjustable weight vectors to produce an output. Matrix inversion and weight vector solutions are feasible only if the matrix equations are linearly independent and the function expansion utilizes appropriate orthogonal polynomials. Among various expansion methods, the triangular expansion stands out due to its relatively low computational complexity and good stability. Unlike the Legendre and Chebyshev expansions, the triangular expansion does not require magnitude normalization [16], making it the preferred method for subsequent studies.

$$\begin{aligned}\boldsymbol{s} = \{&x, \sin[\pi x], \cos[\pi x], \sin[2\pi x], \cos[2\pi x],\\ &\ldots, \sin[(2P+1)\pi x], \cos[(2P+1)\pi x]\}\end{aligned} \tag{1}$$

where $\boldsymbol{s}$ denotes the column vectors of the matrix $\boldsymbol{S}$, the order of the function expansion is $P$, $M = L(4P+3)$ and the length of the input vectors is $N$.

The input-output relationship can be obtained after trigonometric

$$y_n = \sum_{i=1}^{M} w_i s_i \tag{2}$$

Eq. (2) can be expressed as:

$$y_n = \boldsymbol{w}\boldsymbol{s}^T \tag{3}$$

where

$$\begin{aligned}\boldsymbol{s} = [&s_{1,1}, s_{1,2}, \ldots, s_{1,2P+1}, s_{2,1}, s_{2,2}, \ldots s_{2,2P+1}, \ldots,\\ &s_{N,1}, s_{N,2}, \ldots, s_{N,2P+1}]^T\end{aligned} \tag{4}$$

$$\boldsymbol{w} = [w_1, w_2, w_3, \ldots, w_M] \tag{5}$$

$$s_{i,j} = \begin{cases} x_i, & j = 1 \\ \sin(l\pi x_i), & j > 1,\ j \text{ is odd} - \text{number} \\ \cos(l\pi x_i), & j > 1,\ j \text{ is even} - \text{number} \end{cases} \tag{6}$$

where $1 \le l \le P$, the input vector $\boldsymbol{x}$ is:

$$\boldsymbol{x} = [x_n, x_{n-1}, \ldots, x_{n-L+1}]^T \tag{7}$$

Combine with Eq. (6), (7), (8) is expressed by:

$$\begin{aligned}\boldsymbol{s} = \{&x_n, \sin[\pi x_n], \cos[\pi x_n], \ldots,\\ &\sin[(2P+1)\pi x_n], \cos[(2P+1)\pi x_n],\\ &\vdots\\ &x_{n-N+1}, \sin[\pi x_{n-N+1}], \cos[\pi x_{n-N+1}], \ldots,\\ &\sin[(2P+1)\pi x_{n-N+1}], \cos[(2P+1)\pi x_{n-N+1}]\}^T\end{aligned} \tag{8}$$

*2.2. Total Least Square Criterion*

The TLS criterion is widely applied to the EIV model. It achieves unbiased estimation by minimizing the Mean Square Error (MSE) of both the input and output disturbances, resulting in an optimal solution. The optimal solution is

$$(\tilde{\boldsymbol{X}}_n^T + \Delta_n)\boldsymbol{w}(n) = \tilde{\boldsymbol{d}}_n^T + \delta_n \tag{9}$$

where $\delta_n \in \mathbb{R}^{L\times 1}$ and $\Delta_n \in \mathbb{R}^{L\times M}$ are the disturbance matrices for the output and input signal, respectively.

$$\tilde{\boldsymbol{X}}_n = [\tilde{\boldsymbol{x}}_n, \tilde{\boldsymbol{x}}_{n-1}, \ldots, \tilde{\boldsymbol{x}}_{n-L+1}] \tag{10}$$

$$\tilde{\boldsymbol{d}}_n = [\tilde{d}_n, \tilde{d}_{n-1}, \ldots, \tilde{d}_{n-L+1}] \tag{11}$$

The TLS optimal solution can be calculated through Eq. (9) [13]:

$$\boldsymbol{w}_n = -\frac{[\boldsymbol{D}\boldsymbol{v}_n]_{1:M}}{[\boldsymbol{D}\boldsymbol{v}_n]_{M+1}} \tag{12}$$

where the eigenvector $\boldsymbol{v}_n$ corresponding to the smallest eigenvalue of the augmented data covariance matrix $\mathrm{K}_n$.

$$\mathrm{K}_n = \boldsymbol{D}^T \begin{bmatrix} \tilde{\boldsymbol{X}}_n \\ \tilde{\boldsymbol{d}}_n \end{bmatrix} \begin{bmatrix} \tilde{\boldsymbol{X}}_n^T & \tilde{\boldsymbol{d}}_n^T \end{bmatrix} \boldsymbol{D} \tag{13}$$

where $\boldsymbol{D} = diag\left\{1,1,...,1,\theta^{-1/2}\right\}$, the noise mean square ratio of the output and input is represented as $\theta$.

$$J_{GD-TLS}(n) = \Re\left(\mathrm{K}_n, \boldsymbol{D}^{-1}\begin{bmatrix} \boldsymbol{w} \\ -1 \end{bmatrix}\right) = \sum_{n=1}^{N} \frac{\left(\tilde{d}_n - \boldsymbol{w}^T \tilde{\boldsymbol{x}}_n\right)^2}{\|\boldsymbol{w}\|^2 + \theta} \tag{14}$$

Calculate the expectation based on the correlated ergodicity of the input and input-output noise yields:

$$J_{GD-TLS}(n) = \mathrm{E}\left[\frac{\left(\tilde{d}_n - \boldsymbol{w}^T \tilde{\boldsymbol{x}}_n\right)^2}{\|\boldsymbol{w}\|^2 + \theta}\right] = \mathrm{E}\left[\frac{e_n^2}{\|\bar{\boldsymbol{w}}\|^2}\right] \tag{15}$$

where mathematical symbol $E[\bullet]$ stands for the expectation operator, $\bar{\boldsymbol{w}} = [\boldsymbol{w}, -\sqrt{\theta}]^T$ is the corrected vector of augmented and broadened weights, $e_n = \tilde{d}_n - \boldsymbol{w}^T \tilde{\boldsymbol{x}}_n$.

*2.3. Error Entropy*

Entropy quantifies the degree of uncertainty associated with variables [14]. For a random variable, the Renyi entropy of order $\zeta(\zeta \neq 0, \zeta > 0)$ is

$$\mathrm{Z}_\zeta(e) = \frac{1}{1-\zeta}\log\int z^\zeta(e)de = \frac{1}{1-\alpha}E[z^{\zeta-1}(e)] \tag{16}$$

where $z(\bullet)$ represents the probability density function (PDF) with respect to $e$. $z(e)$ is typically evaluated using the Parzen window strategy based on the given error samples:

$$z_\zeta(e) \approx \hat{z}_\zeta(e) = \frac{1}{N}\sum_{i=1}^{N} G_\sigma(e - e_i) \tag{17}$$

where in the kernel function $G_\sigma(e) = \frac{1}{\sqrt{2\pi}\sigma}\exp(-\frac{e^2}{2\sigma^2})$, $\sigma > 0$ is kernel bandwidth, and $\{e_i\}_{i=1}^N$ indicates the error set with $N$ error samples. As the error probability distribution of random variables is indeterminable, Renyi's quadratic entropy is evaluated using the Parzen window method, as described in [17], the estimation of $\mathrm{Z}_\zeta(e)$ can be obtained.

$$\mathrm{Z}_\zeta(e) \approx \hat{\mathrm{Z}}_2(e) = -\log\frac{1}{L^2}\sum_{i=1}^{N}\sum_{j=1}^{N} G_\sigma(e_i - e_j) \tag{18}$$

## 3. Proposed Algorithms

Within this section, we introduce the MEE and TLS criteria to handle the non-Gaussian input-output noise. Given the triangular expansion's low computational complexity and high stability, a FLANN is used to model the nonlinear system. Based on this, in Part. 3.1, MTEE-FLANN algorithm is proposed. Then, to reduce the computational complexity, a quantized method is utilized, QMTEE-FLANN algorithm is introduced in Part. 3.2.

*3.1. Error Entropy*

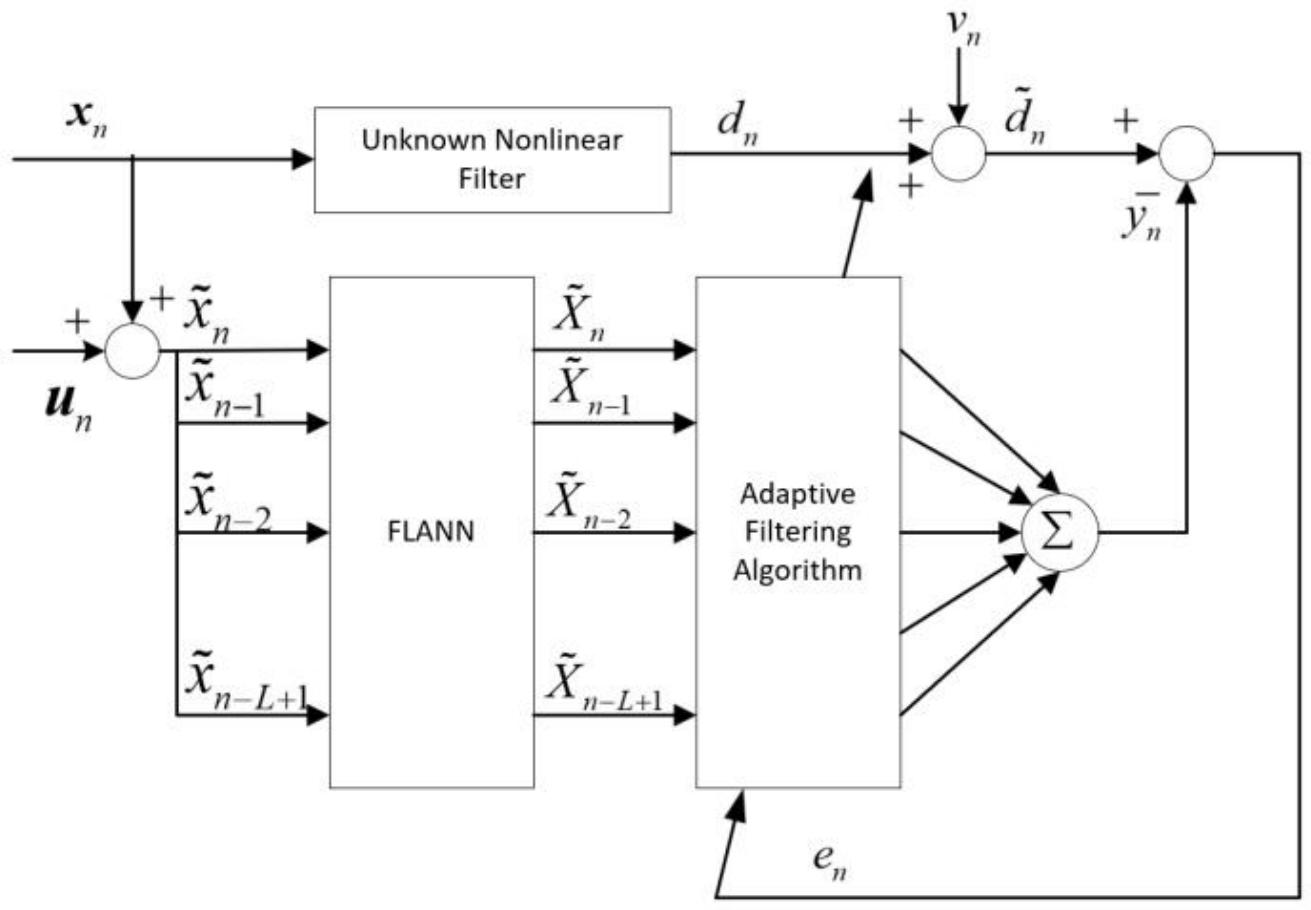

Figure. 2. Nonlinear system identification model under the EIV model.

Correntropy is given two random variables $X$ and $Y$. In [18], the definition of correntropy is

$$\tilde{\boldsymbol{x}}_n = \boldsymbol{x}_n + \boldsymbol{u}_n \tag{19}$$

$$\tilde{d}_n = d_n + v_n \tag{20}$$

where the self-covariance matrix of the input noise vector $\boldsymbol{u}_n \in \mathbb{R}^{L\times 1}$ is $\sigma_u^2 I$ , $v_n \in \mathbb{R}$ is the output noise signal and the self-covariance matrix of $v_n$ is $\sigma_v^2 I$ .

Assuming $\tilde{\boldsymbol{x}}_n \in \mathbb{R}^{L\times 1}, \tilde{\boldsymbol{x}}_n = [\tilde{x}_n, \tilde{x}_{n-1}, \ldots, \tilde{x}_{n-L+1}]^T$ is an external input vector to the system, FLANN is a triangular function expansion of finite order *P*. The function expansion is:

$$\begin{aligned} \tilde{\boldsymbol{X}}_n = \{&\tilde{X}_n, \tilde{X}_{n-1}, \ldots, \tilde{X}_{n-L+1}\}^T \\ &\{\tilde{x}_n, \sin(\pi\tilde{x}_n), \cos(\pi\tilde{x}_n), \ldots, \\ &\sin[(2P+1)\pi\tilde{x}_n], \cos[(2P+1)\pi\tilde{x}_n] \\ &\vdots \\ &\tilde{x}_{n-L+1}, \sin(\pi\tilde{x}_{n-L+1}), \cos(\pi\tilde{x}_{n-L+1}), \ldots, \\ &\sin[(2P+1)\pi\tilde{x}_{n-L+1}], \cos[(2P+1)\pi\tilde{x}_{n-L+1}]\}^T \end{aligned} \tag{21}$$

There are the following nonlinear systems of *L*-dimension:

$$d_n = f(\boldsymbol{x}_n) = \boldsymbol{h}^T \boldsymbol{X}_n \tag{22}$$

where $\boldsymbol{h} \in \mathbb{R}^{L(4P+3)\times 1}$ is the optimal filter coefficients obtained by the FLANN filter, which can be computed by the above equation. The unknown system vector needs to be evaluated. $\boldsymbol{X}_n \in \mathbb{R}^{L(4P+3)\times 1}$ and $d_n \in \mathbb{R}$ represent the input signal expanded by noise-free functions and the accurately observed output signal, respectively. The error signal is

$$e_n = \tilde{d}_n - y_n = \tilde{d}_n - \boldsymbol{W}_n^T \tilde{\boldsymbol{X}}_n \tag{23}$$

The input noise is:

$$
\begin{aligned}
\boldsymbol{U}_n &= \tilde{\boldsymbol{X}}_n - \boldsymbol{X}_n \\
&= \{u_n, \sin(\pi\tilde{x}_n) - \sin(\pi x_n), \\
&\quad \cos(\pi\tilde{x}_n) - \cos(\pi x_n), \ldots, \\
&\quad \sin[(2P+1)\pi\tilde{x}_n] - \sin[(2P+1)\pi x_n], \\
&\quad \cos[(2P+1)\pi\tilde{x}_n] - \cos[(2P+1)\pi x_n], \\
&\quad \vdots \\
&\quad u_{n-L+1}, \sin(\pi\tilde{x}_{n-L+1}) - \sin(\pi x_{n-L+1}), \\
&\quad \cos(\pi\tilde{x}_{n-L+1}) - \cos(\pi x_{n-L+1}), \ldots, \\
&\quad \sin[(2P+1)\pi\tilde{x}_{n-L+1}] - \sin[(2P+1)\pi x_{n-L+1}], \\
&\quad \cos[(2P+1)\pi\tilde{x}_{n-L+1}] - \cos[(2P+1)\pi x_{n-L+1}]\}^T
\end{aligned} \tag{24}
$$

$$
\begin{aligned}
\boldsymbol{E}[\boldsymbol{U}_n^2] \approx &\frac{1}{L(4P+3)} E[u_n^2 + \pi^4 u_n^2 \cos^2(\pi u_n) + \\
&\pi^4 u_n^2 \sin^2(\pi u_n) + \ldots + \\
&(2\text{P}+1)^4 \pi^4 u_n^2 \cos^2[(2P+1)\pi u_n] + \\
&(2\text{P}+1)^4 \pi^4 u_n^2 \sin^2[(2P+1)\pi u_n] + \\
&\vdots \\
&u_{n-L+1}^2 + \pi^4 u_{n-L+1}^2 \cos^2(\pi u_{n-L+1}) + \\
&\pi^4 u_{n-L+1}^2 \sin^2(\pi u_{n-L+1}) + \ldots + \\
&(2\text{P}+1)^4 \pi^4 u_{n-L+1}^2 \cos^2[(2P+1)\pi u_{n-L+1}] + \\
&(2\text{P}+1)^4 \pi^2 u_{n-L+1}^2 \sin^2[(2P+1)\pi u_{n-L+1}]] \\
\approx &\sigma_U^2 = \varphi\sigma_u^2
\end{aligned} \tag{25}
$$

Compared to the input signal, the input noise is very small. Use the Taylor expansion, Eq. (25), (26) can be obtained.

$$
\varphi = \frac{1 + \pi^4 P\{1 + (P+1)\left[\frac{16}{5}(P+2)(P+3)(P+4) - 16P(P+1) - \frac{76}{3}(2P+1) - 44\right]\}}{4P+3} \tag{26}
$$

Next, using Eq. (18), the residual signal entropy minimization problem is equal to the Quadratic Information Potential (Q-IP) maximization problem

$$
\arg\min_{w} \mathrm{Z}_2(e_n) = arg\max_{w} QIP_2^{MEE}(e_n) \tag{27}
$$

where $QIP_2^{MEE}(e_n) = \frac{1}{N^2}\sum_{i=1}^{N}\sum_{j=1}^{N} G_\sigma(e_i - e_j)$.

Therefore, combine with the TLS criterion, the cost function of MTEE criterion is

$$
\begin{aligned}
J_{MTEE} &= \frac{1}{N^2}\sum_{i=1}^{N}\sum_{j=1}^{N} G_\sigma\Big(\frac{e_i - e_j}{\sqrt{\|\boldsymbol{w}\|^2 + \theta}}\Big) \\
&= \frac{1}{N^2}\sum_{i=1}^{N}\sum_{j=1}^{N} G_\sigma\Big(\frac{e_i - e_j}{\|\bar{\boldsymbol{w}}\|}\Big)
\end{aligned} \tag{28}
$$

In the FLANN model, to minimise the perturbations of the input and output, the MTEE method is as

$$
\begin{aligned}
J_{MTEE-FLANN}(\boldsymbol{W}_n) &= \frac{1}{N^2}\sum_{i=1}^{N}\sum_{j=1}^{N} G_\sigma\Big(\frac{e_i - e_j}{\sqrt{\|\boldsymbol{W}_n\|^2 + \gamma}}\Big) \\
&= \frac{1}{N^2}\sum_{i=1}^{N}\sum_{j=1}^{N} G_\sigma\Big(\frac{e_i - e_j}{\|\bar{\boldsymbol{W}}_n\|}\Big)
\end{aligned} \tag{29}
$$

where $\bar{W}_n = [\sqrt{\gamma}, -W_n]^T$ is the modified augmented weight vector, $\gamma = \sigma_v^2 / \sigma_U^2$ .

Using gradient descent and considering instantaneous values instead of expectations, the weight update formula for the MTEE-FLANN algorithm is:

$$W_{n+1} = W_n + \frac{\mu}{\sigma^2 N^2}\left[\sum_{i=1}^{N}\sum_{j=1}^{N} G_\sigma\left(\frac{\Delta e_{ij}}{\|\bar{W}_n\|}\right)\frac{\|\bar{W}_n\|^2 \Delta e_{ij}\Delta\tilde{X}_{ij} + \Delta e_{ij}^2 W_n}{\|\bar{W}_n\|^4}\right] \tag{30}$$

where $\mu$ is the step parameter and $\Delta a_{ij} = a_i - a_j, (i > 0, j > 0)$ .

In an online scenario, at time instant $i$ , the latest $M$ data points $\{(e_j) | j = 1 - M + 1, \ldots, i\}$ is utilized to evaluate the QIP based on the $L$ samples. To further optimal computational cost, the QIP evaluate is approximated stochastically by omitting the time average in Eq. (29), allowing Eq. (30) to be reformulated.

$$W_{n+1} = W_n + \frac{\mu}{\sigma^2 M}\left[\sum_{j=i-M+1}^{i} G_\sigma\left(\frac{\Delta e_{ij}}{\|\bar{W}_n\|}\right)\frac{\|\bar{W}_n\|^2 \Delta e_{ij}\Delta\tilde{X}_{ij} + \Delta e_{ij}^2 W_n}{\|\bar{W}_n\|^4}\right] \tag{31}$$

Table 1 The Iterative Process of the MTEE-FLANN Algorithm.

**Initialization:** $W_1 = \mathbf{0}$

**Parameters:** $\mu, P, \sigma, \gamma = \sigma_v^2 / \sigma_U^2, \boldsymbol{h}$

**Input: data sequences** $\{d(n), x_n, u_n, v_n\}, n = 1, 2, \ldots$

**For** $n$= 0,1,2, ...

$\tilde{x}_n = x_n + u_n$

$d_n = h^T X_n$

$\tilde{d}_n = d_n + v_n$

$\tilde{X}_n = \{\tilde{X}_n, \tilde{X}_{n-1}, \ldots, \tilde{X}_{n-L+1}\}^T$

$\{\tilde{x}_n, \sin(\pi\tilde{x}_n), \cos(\pi\tilde{x}_n), \ldots,$

$\sin[(2P+1)\pi\tilde{x}_n], \cos[(2P+1)\pi\tilde{x}_n]$

$\vdots$

$\tilde{x}_{n-L+1}, \sin(\pi\tilde{x}_{n-L+1}), \cos(\pi\tilde{x}_{n-L+1}), \ldots,$

$\sin[(2P+1)\pi\tilde{x}_{n-L+1}], \cos[(2P+1)\pi\tilde{x}_{n-L+1}]\}^T$

$e_n = \tilde{d}_n - W_n^T \tilde{X}_n$

$\bar{W}_n = [\sqrt{\gamma}, -W_n]^T$

$J_{MTEE-FLANN}(W_n) = \frac{1}{N^2}\sum_{i=1}^{N}\sum_{j=1}^{N} G_\sigma\left(\frac{e_i - e_j}{\|\bar{W}_n\|}\right)$

$W_{n+1} = W_n + \frac{\mu}{\sigma^2 M}\left[\sum_{j=i-M+1}^{i} G_\sigma\left(\frac{\Delta e_{ij}}{\|\bar{W}_n\|}\right)\frac{\|\bar{W}_n\|^2 \Delta e_{ij}\Delta\tilde{X}_{ij} + \Delta e_{ij}^2 W_n}{\|\bar{W}_n\|^4}\right]$

**End**

### *3.2. The Design of QMTEE-FLANN Algorithm*

The high computational complexity of MEE-type algorithms makes them unsuitable for real-time applications. To address this issue, a quantization strategy [15] is employed, significantly reducing the computational cost of the MTEE-FLANN. This leads to the development of the QMTEE-FLANN algorithm.

The primary factor for the high complexity of the MTEE-FLANN algorithm lies in the operations involved in the MEE component. To optimal the double summation process within the MEE part and obtain an error distribution close to the MEE criterion, we focus on keeping the errors close to their mean. As a result, the cost function for the QMTEE criterion is formulated as:

$$J_{QMTEE} = \frac{1}{L^2}\sum_{i=1}^{M} G_\sigma\left(\frac{e_i - \overline{e}_i}{\|\overline{\boldsymbol{w}}\|}\right) \tag{32}$$

where $\overline{e}_i$ indicates the error mean within sample window

$$\overline{e}_i = \frac{1}{M}\sum_{j=i-M}^{i-1} e_j \tag{33}$$

This approach utilizes existing error estimates for average error calculation, which simplifies the calculation process and thus improves the efficiency and effectiveness of the algorithm implementation.

As a result, the weight update formula for the QMTEE-FLANN is:

$$\boldsymbol{W}_{n+1} = \boldsymbol{W}_n + \frac{\mu}{\sigma^2 M}\left[\sum_{j=i-M+1}^{i} G_\sigma\left(\frac{e_i - \overline{e}_i}{\|\overline{\boldsymbol{W}}_n\|}\right)\frac{\left[\|\overline{\boldsymbol{W}}_n\|^2 (e_i - \overline{e}_i)(\tilde{\boldsymbol{X}}_i - \overline{\tilde{\boldsymbol{X}}}_i) + (e_i - \overline{e}_i)^2 \boldsymbol{W}_n\right]}{\|\overline{\boldsymbol{W}}_n\|^4}\right] \tag{34}$$

where $\overline{\tilde{\boldsymbol{X}}}_i$ is

$$\overline{\tilde{\boldsymbol{X}}}_i = \frac{1}{L}\sum_{j=i-M}^{i-1} \tilde{\boldsymbol{X}}_i \tag{35}$$

## 4. Performance Discussion

This section focuses on the local stability of MTEE-FLANN in the neighborhood of the desired value of the parameter vector $\boldsymbol{h}$. Under certain conditions, $\boldsymbol{h}$ is approximately a local extremum of Eq. (29), under some specific conditions. For the purpose of this analysis, several assumptions are made below.

*Assumption* 1: The input signal $\boldsymbol{X}_n$ is independent and identically distributed, and its covariance matrix $\boldsymbol{R}_X = \boldsymbol{E}[\boldsymbol{X}_n \boldsymbol{X}_n^T]$ is positive definite and full rank [19].

*Assumption* 2: The input and output noise, $\boldsymbol{U}_n$ and $v_n$, are zero-mean Gaussian noise. Additionally, $\boldsymbol{U}_n$, $v_n$ and the weight vector $\boldsymbol{W}_n$ are independent of $\boldsymbol{X}_n$ [19][20].

### *4.1. Local Extreme point of MTEE-FLANN*

Define Eq. (32)

$$\begin{cases} e_n^{TOT} = \dfrac{\tilde{d}_n - \boldsymbol{W}_n^T \tilde{\boldsymbol{X}}_n}{\|\overline{\boldsymbol{W}}_n\|} = \dfrac{v_n - \boldsymbol{W}_n^T \boldsymbol{U}_n}{\|\overline{\boldsymbol{W}}_n\|} \\ e_n^{TOT(\boldsymbol{h})} = \dfrac{\tilde{d}_n - \boldsymbol{h}^T \tilde{\boldsymbol{X}}_n}{\|\overline{\boldsymbol{h}}\|} = \dfrac{v_n - \boldsymbol{h}^T \boldsymbol{U}_n}{\|\overline{\boldsymbol{h}}\|} \\ \rho_n = e_n^{TOT}\|\overline{\boldsymbol{h}}\| = v_n - \boldsymbol{h}^T \boldsymbol{U}_n \end{cases} \tag{36}$$

Substituting $\boldsymbol{h}$ into the the gradient of Eq. (29)

$$\nabla J_{MTEE-FLANN}(\boldsymbol{h}) = \frac{1}{\sigma^2 L^2}\sum_{i=1}^{M}\sum_{j=1}^{M} G_\sigma(\Delta e_n^{TOT(\boldsymbol{h})})\left(\frac{\Delta\rho_{ij}\Delta\tilde{\boldsymbol{X}}_{ij}}{\|\overline{\boldsymbol{h}}\|^2} + \frac{\Delta\rho_{ij}^2 \boldsymbol{h}}{\|\overline{\boldsymbol{h}}\|^4}\right) \tag{37}$$

The Gaussian term in Eq. (37) satisfies $0 < G_\sigma(\Delta e_n^{TOT(\boldsymbol{h})}) \le 1$. The total error should generally be small in the case of perfect estimation when input and output signal-to-noise ratio (SNR) are not too low, . Consequently, for a relatively large kernel size $\sigma$, the value $\Delta e_n^{TOT(\boldsymbol{h})}/\sigma$ of is generally close to zero (See next page for detailed analysis). As a result, the Gaussian term approaches one. By applying the approximation and taking the absolute value of both sides of Eq. (38), we obtain:

$$\left|\nabla J_{MTEE-FLANN}(\boldsymbol{h})\right| \approx \left|\frac{1}{\sigma^2 M^2}\sum_{i=1}^{M}\sum_{j=1}^{M}\left(\frac{\Delta\rho_{ij}\Delta\tilde{\boldsymbol{X}}_{ij}}{\|\overline{\boldsymbol{h}}\|^2} + \frac{\Delta\rho_{ij}^2 \boldsymbol{h}}{\|\overline{\boldsymbol{h}}\|^4}\right)\right| \tag{38}$$

Since $\boldsymbol{X}_n$, as well as $\boldsymbol{U}_n$ and $v_n$, are generated by stationary and ergodic processes, the double time average in Eq. (38) is substituted with the ensemble expectation, resulting in:

$$\left|\nabla J_{MTEE-FLANN}(\boldsymbol{h})\right| \approx \left|\nabla \hat{J}_{MTEE-FLANN}(\boldsymbol{h})\right| = \left|\frac{1}{\sigma^2}E[\frac{\triangle\rho_{ij}\triangle\tilde{\boldsymbol{X}}_{ij}}{\left\|\bar{\boldsymbol{h}}\right\|^2}+\frac{\triangle\rho_{ij}^2\boldsymbol{h}}{\left\|\bar{\boldsymbol{h}}\right\|^4}]\right| \tag{39}$$

According to Assumption 1 and 2, the second term in the right hand of Eq. (39) is represented as

$$\begin{aligned}\frac{1}{\sigma^2}E[\frac{\triangle\rho_{ij}\triangle\tilde{\boldsymbol{X}}_{ij}}{\left\|\bar{\boldsymbol{h}}\right\|^2}] &= \frac{1}{\left\|\bar{\boldsymbol{h}}\right\|^2\sigma^2}E[(\rho_i-\rho_j)(\tilde{\boldsymbol{X}}_i-\tilde{\boldsymbol{X}}_j)] \\ &= -\frac{1}{\left\|\bar{\boldsymbol{h}}\right\|^2\sigma^2}E[\boldsymbol{U}_i^T\boldsymbol{h}\boldsymbol{U}_i+\boldsymbol{U}_j^T\boldsymbol{h}\boldsymbol{U}_j] \\ &= -\frac{2\sigma_U^2}{\left\|\bar{\boldsymbol{h}}\right\|^2\sigma^2}\boldsymbol{h}\end{aligned} \tag{40}$$

To calculate the first term in the right hand of Eq. (40), (41) and (42) need to be calculated.

$$\begin{aligned}E[\rho_i^2] &= E[(v_i-\boldsymbol{h}^T\boldsymbol{U}_i)^2] \\ &= E[v_i^2]+\boldsymbol{h}^TE[\boldsymbol{U}_i\boldsymbol{U}_i^T]\boldsymbol{h} \\ &= \sigma_v^2+(\left\|\bar{\boldsymbol{h}}\right\|^2-\gamma)\sigma_U^2\end{aligned} \tag{41}$$

$$\begin{aligned}E[\triangle\rho_{ij}^2] &= E[\rho_i^2+\rho_j^2-2\rho_i\rho_j] \\ &= 2E[\rho_i^2] \\ &= 2[\sigma_v^2+(\left\|\bar{\boldsymbol{h}}\right\|^2-\gamma)\sigma_U^2]\end{aligned} \tag{42}$$

Therefore,

$$\begin{aligned}\frac{1}{\sigma^2}E[\frac{\triangle\rho_{ij}^2\boldsymbol{h}}{\left\|\bar{\boldsymbol{h}}\right\|^4}] &= \frac{1}{\left\|\bar{\boldsymbol{h}}\right\|^2}E[\frac{\triangle\rho_{ij}^2}{\left\|\bar{\boldsymbol{h}}\right\|^2\sigma^2}]\boldsymbol{h} \\ &= \frac{2}{\left\|\bar{\boldsymbol{h}}\right\|^2}\frac{[\sigma_v^2+(\left\|\bar{\boldsymbol{h}}\right\|^2-\gamma)\sigma_U^2]}{\left\|\bar{\boldsymbol{h}}\right\|^2\sigma^2}\boldsymbol{h}\end{aligned} \tag{43}$$

Eq. (39) can be rewrite as

$$\begin{aligned}\left|\nabla\hat{J}_{MTEE-FLANN}(\boldsymbol{h})\right| &= \left|-\frac{2\sigma_U^2}{\left\|\bar{\boldsymbol{h}}\right\|^2\sigma^2}\boldsymbol{h}+\frac{2}{\left\|\bar{\boldsymbol{h}}\right\|^2}\frac{[\sigma_v^2+(\left\|\bar{\boldsymbol{h}}\right\|^2-\gamma)\sigma_U^2]}{\left\|\bar{\boldsymbol{h}}\right\|^2\sigma^2}\boldsymbol{h}\right| \\ &= \left|\frac{2[\sigma_v^2-\gamma\sigma_U^2]}{\left\|\bar{\boldsymbol{h}}\right\|^4\sigma^2}\boldsymbol{h}\right|\end{aligned} \tag{44}$$

Thus, the gradient above is zero, denoting that $\boldsymbol{h}$ is a local extremum of the cost function when the noise intensity is known as prior knowledge ( $\gamma=\sigma_v^2/\sigma_U^2$ ),.

In less optimistic cases, when the noise intensity is unknown and $\lambda$ approaches one, Eq. (45) can be obtained.

$$0<\left|\frac{2[\sigma_v^2-\sigma_U^2]}{\left\|\bar{\boldsymbol{h}}\right\|^4\sigma^2}\boldsymbol{h}\right|<\left|\frac{[\sigma_v^2-\sigma_U^2]}{\sigma^2}\right|\boldsymbol{R} \tag{45}$$

where $\boldsymbol{R}$ is a $L(4P+3)\times 1$ vector, all elements are equivalent to one. Eq. (45) follows from $\left|\frac{\boldsymbol{h}}{\left\|\boldsymbol{h}\right\|^2+1}\right|<\frac{1}{2}\boldsymbol{R}$ and $\left\|\bar{\boldsymbol{h}}\right\|^2>\gamma=1$. Thus when $\sigma_v^2$ is comparable to $\sigma_U^2$ and with a relatively large kernel size $\sigma^2$, $\left|\frac{[\sigma_v^2-\sigma_U^2]}{\sigma^2}\right|$ would be

close to zero. Therefore, $\left|\nabla\hat{J}_{MTEE-FLANN}(\boldsymbol{h})\right|$ is significantly smaller than $\boldsymbol{h}$ but also close to zero, suggesting that $\boldsymbol{h}$ approaches a local extremum of the cost function [21].

In addition, to further illustrate the using of $G_\sigma(\triangle e_n^{TOT(\boldsymbol{h})}) \approx 1$ , $\chi_{ij} = \triangle e_n^{TOT(\boldsymbol{h})} / \sigma$ is close to zero in general situations by proving $E[\chi_{ij}^2] = E[\frac{\triangle\rho_{ij}^2}{\left\|\bar{\boldsymbol{h}}\right\|^2 \sigma^2}] \approx 0$ . For simplicity, the auto-covariance matrix of input is assumed as $\sigma_X^2 \boldsymbol{I}$, where $\sigma_X^2$ represents the signal intensity. The input SNR is then indicated as:

$$SNR_U = 10\log\frac{\sigma_X^2}{\sigma_U^2} \tag{46}$$

$$SNR_v = 10\log\frac{(\left\|\bar{\boldsymbol{h}}\right\|^2 - \gamma)\sigma_X^2}{\sigma_v^2} \tag{47}$$

Based on these definitions, $E[\frac{\triangle\rho_{ij}^2}{\left\|\bar{\boldsymbol{h}}\right\|^2 \sigma^2}]$ can be rewrite as

$$\begin{aligned} E[\chi_{ij}^2] &= 2\frac{\sigma_v^2 + (\left\|\bar{\boldsymbol{h}}\right\|^2 - \gamma)\sigma_U^2}{\left\|\bar{\boldsymbol{h}}\right\|^2 \sigma^2} \\ &= \frac{2(\left\|\bar{\boldsymbol{h}}\right\|^2 - \gamma)}{\left\|\bar{\boldsymbol{h}}\right\|^2}\frac{\sigma_X^2}{\sigma^2}[\frac{\sigma_v^2}{(\left\|\bar{\boldsymbol{h}}\right\|^2 - \gamma)\sigma_X^2} + \frac{\sigma_U^2}{\sigma_X^2}] \\ &= \frac{2(\left\|\bar{\boldsymbol{h}}\right\|^2 - \gamma)}{\left\|\bar{\boldsymbol{h}}\right\|^2}\frac{\sigma_X^2}{\sigma^2}(10^{-SNR_v/10} + 10^{-SNR_U/10}) \end{aligned} \tag{48}$$

Thus, when the input and output SNR are sufficiently high, (e.g., 10 ~ 20 dB),

$$E[\chi_{ij}^2] \ll \frac{\sigma_X^2}{\sigma^2} \tag{49}$$

If the square of $\sigma$ is equivalent to the signal intensity, the ratio on the right-hand side of Eq. (49) approaches one, and $E[\chi_{ij}^2]$ approaches zero. Furthermore, in high-SNR scenarios, even if $\sigma_X^2/\sigma^2$ is moderately larger than one, $E[\chi_{ij}^2]$ approaches zero. This result confirms that $\chi_{ij}$ is actually approaching zero overall, making it reasonable to use the approximation $G_\sigma(\triangle e_n^{TOT(\boldsymbol{h})}) \approx 1$ . In conclusion, $\boldsymbol{h}$ is exactly or approximately a local extremum of the cost function for the MTEE-FLANN under specific conditions.

*4.2. Local Stability of MTEE-FLANN*

Assuming that the approximate value of the parameter vector resides in the vicinity of $\boldsymbol{h}$ when the algorithm gets nearly converged. Subtract both sides of Eq. (30) from $\boldsymbol{h}$

$$\tilde{\boldsymbol{W}}_{n+1} = \tilde{\boldsymbol{W}}_n - \mu\nabla\hat{J}_{MTEE-FLANN}(\boldsymbol{W}_n) \tag{50}$$

where $\tilde{\boldsymbol{W}}_n = \boldsymbol{h} - \boldsymbol{W}_n$ .

Based on [22], it is assumed that $\hat{J}_{MTEE-FLANN}(\boldsymbol{W})$ is twice continuously differentiable within the vicinity of the line segment joining the points $\boldsymbol{h}$ and $\tilde{\boldsymbol{W}}_n = \boldsymbol{h} - \boldsymbol{W}_n$ , Eq. (51) is obtained

$$
\begin{aligned}
&\nabla \hat{J}_{MTEE-FLANN}(\boldsymbol{W}_n) \\
&= \nabla \hat{J}_{MTEE-FLANN}(\boldsymbol{h}) - [\int_0^1 H_{MTEE-FLANN}(\boldsymbol{h} - t\tilde{\boldsymbol{W}}_n)dt]\tilde{\boldsymbol{W}}_n \\
&\approx -[\int_0^1 H_{MTEE-FLANN}(\boldsymbol{h} - t\tilde{\boldsymbol{W}}_n)dt]\tilde{\boldsymbol{W}}_n \\
&\approx -[\int_0^1 H_{MTEE-FLANN}(\boldsymbol{h})dt]\tilde{\boldsymbol{W}}_n \\
&\approx -H_{MTEE-FLANN}(\boldsymbol{h})\tilde{\boldsymbol{W}}_n
\end{aligned} \tag{51}
$$

where $H_{MTEE-FLANN}(\boldsymbol{h})$ is the Hessian matrix of $J_{MTEE-FLANN}(\boldsymbol{h})$. The second approximation is based on the assumption that $\boldsymbol{W}_n$ lies in the neighborhood of $\boldsymbol{h}$, making $\tilde{\boldsymbol{W}}_n$ relatively small. Substituting this result into Eq. (50) allows Eq. (52) to be computed.

$$
\tilde{\boldsymbol{W}}_{n+1} \approx [\boldsymbol{I} + \mu H_{MTEE-FLANN}(\boldsymbol{h})]\tilde{\boldsymbol{W}}_n \tag{52}
$$

It is essential that all eigenvalues of $I + \mu H_{MTEE-FLANN}(\boldsymbol{h})$ must have magnitudes smaller than one [18] [23] in order to guarantee local stability. Therefore, we have

$$
|1 + \mu \varpi_r| < 1, r = 1, \ldots, R \tag{53}
$$

where $\{\varpi_r | r = 1, \ldots, R\}$ are the eigenvalues of $H_{MTEE-FLANN}(\boldsymbol{h})$.

Then, in order to calculate $H_{MTEE-FLANN}(\boldsymbol{h})$, calculate the gradient vector $\triangle e_{ij}^{TOT}$ at $\boldsymbol{h}$ is the first step.

$$
\boldsymbol{q} = \frac{\partial \triangle e_{ij}^{TOT}}{\partial \boldsymbol{W}_n}\Big|_{\boldsymbol{h}} = -\frac{\triangle p_{ij}\boldsymbol{h}}{\|\bar{\boldsymbol{h}}\|^3} + \frac{\triangle \tilde{\boldsymbol{X}}_{ij}}{\|\bar{\boldsymbol{h}}\|} \tag{54}
$$

And calculating dual gradient of $\triangle e_{ij}^{TOT}$ at $\boldsymbol{h}$.

$$
\begin{aligned}
\boldsymbol{Q} &= \frac{\partial^2 \triangle e_{ij}^{TOT}}{\partial \boldsymbol{W}_n^2}\Big|_{\boldsymbol{h}} \\
&= -\frac{\boldsymbol{h}\triangle \tilde{\boldsymbol{X}}_{ij}^T}{\|\bar{\boldsymbol{h}}\|^3} + \triangle p_{ij}\frac{\|\bar{\boldsymbol{h}}\|\boldsymbol{I} - 2\boldsymbol{H}}{\|\bar{\boldsymbol{h}}\|^5} - \frac{\boldsymbol{h}\boldsymbol{q}^T}{\|\bar{\boldsymbol{h}}\|} \\
&= -\frac{2\boldsymbol{h}\triangle \tilde{\boldsymbol{X}}_{ij}^T}{\|\bar{\boldsymbol{h}}\|^3} - \frac{3\triangle p_{ij}\boldsymbol{H}}{\|\bar{\boldsymbol{h}}\|^5} + \frac{\triangle p_{ij}\boldsymbol{I}}{\|\bar{\boldsymbol{h}}\|^3}
\end{aligned} \tag{55}
$$

where $\boldsymbol{H} = \boldsymbol{h}\boldsymbol{h}^T$.

Then, $\tilde{\boldsymbol{\Phi}}$ is defined as

$$
\begin{aligned}
\tilde{\boldsymbol{\Phi}} &= E[\tilde{\boldsymbol{X}}_{ij}\tilde{\boldsymbol{X}}_{ij}^T] \\
&= E[\boldsymbol{X}_{ij}\boldsymbol{X}_{ij}^T] + E[\boldsymbol{U}_{ij}\boldsymbol{U}_{ij}^T] \\
&= (\sigma_X^2 + \sigma_U^2)\boldsymbol{I}
\end{aligned} \tag{56}
$$

Therefore, $E[\triangle \tilde{\boldsymbol{X}}_{ij} \triangle \tilde{\boldsymbol{X}}_{ij}^T] = 2\tilde{\boldsymbol{\Phi}}$.

Based on the above and $G_\sigma(\triangle e_{ij}^{TOT}) \approx 1$, $H_{MTEE-FLANN}(\boldsymbol{h})$ is calculated as Eq. (57), where $\chi_{ij} = \triangle e_n^{TOT(\boldsymbol{h})} / \sigma$ is generally smaller than one and close to zero so that the part of $\chi_{ij}^2$ can be omitted and the approximately equal sign holds.

According to Eq. (57), it can be shown that the absolute value of each element in $\boldsymbol{H} / \|\bar{\boldsymbol{h}}\|^2$ is less than one can be proved. Thus, $\hat{H}_{MTEE-FLANN}(\boldsymbol{h})$ is predominantly influenced by the term $-\frac{2(\|\bar{\boldsymbol{h}}\|^2 \sigma_X^2 + \sigma_U^2 - \sigma_v^2)}{\|\bar{\boldsymbol{h}}\|^4 \sigma^2}\boldsymbol{I}$ when the input and output SNR are not too low, particularly, when the input noise intensity is approximately equal to the output noise intensity. Hence, a practical estimation of $\hat{H}_{MTEE-FLANN}(\boldsymbol{h})$'s eigenvalue, i.e., Eq. (58), can be obtained.

A reasonable estimation of $\hat{H}_{MTEE-FLANN}(\boldsymbol{h})$'s eigenvalue can be derived, as given in Eq. (58).

$$\begin{aligned}
&\hat{H}_{MTEE-FLANN}(\boldsymbol{h}) \\
&= \frac{\partial \nabla \hat{J}_{MTEE-FLANN}(\boldsymbol{W}_n)}{\partial \boldsymbol{W}_n}\Big|_{\boldsymbol{h}} \\
&= \frac{1}{\sigma^2} E[\frac{(\triangle e_{ij}^{TOT(\boldsymbol{h})})^2 \boldsymbol{q}\boldsymbol{q}^T}{\sigma^2} - \boldsymbol{q}\boldsymbol{q}^T + \triangle e_{ij}^{TOT(\boldsymbol{h})}\boldsymbol{Q}] \\
&= \frac{1}{\sigma^2} E[(\chi_{ij}^2 - 1)\frac{\triangle \tilde{\boldsymbol{X}}_{ij} \triangle \tilde{\boldsymbol{X}}_{ij}^T}{\|\bar{\boldsymbol{h}}\|^2} + (\chi_{ij}^2 - 1)\frac{\triangle \rho_{ij}(\boldsymbol{h}\triangle \tilde{\boldsymbol{X}}_{ij}^T)^T}{\|\bar{\boldsymbol{h}}\|^4} \\
&\quad + (\chi_{ij}^2 - 3)\frac{\triangle \rho_{ij}\boldsymbol{h}\triangle \tilde{\boldsymbol{X}}_{ij}^T}{\|\bar{\boldsymbol{h}}\|^4} + (\chi_{ij}^2 - 4)\frac{\triangle \rho_{ij}^2}{\|\bar{\boldsymbol{h}}\|^6}\boldsymbol{H} + \frac{\triangle \rho_{ij}^2}{\|\bar{\boldsymbol{h}}\|^4}\boldsymbol{I}] \\
&\approx -\frac{2\tilde{\boldsymbol{\Phi}}}{\|\bar{\boldsymbol{h}}\|^2 \sigma^2} - \frac{E[\triangle \rho_{ij}(\boldsymbol{h}\triangle \tilde{\boldsymbol{X}}_{ij}^T)^T]}{\|\bar{\boldsymbol{h}}\|^4 \sigma^2} - \frac{3E[\triangle \rho_{ij}\boldsymbol{h}\triangle \tilde{\boldsymbol{X}}_{ij}^T]}{\|\bar{\boldsymbol{h}}\|^4 \sigma^2} - 4\frac{E[\chi_{ij}^2]}{\|\bar{\boldsymbol{h}}\|^4}\boldsymbol{H} + \frac{E[\chi_{ij}^2]}{\|\bar{\boldsymbol{h}}\|^2}\boldsymbol{I} \\
&\approx -\frac{2(\sigma_X^2 + \sigma_U^2)}{\|\bar{\boldsymbol{h}}\|^2 \sigma^2}\boldsymbol{I} + \frac{8\sigma_U^2 \boldsymbol{H}}{\|\bar{\boldsymbol{h}}\|^4 \sigma^2} - \frac{4E[\chi_{ij}^2]}{\|\bar{\boldsymbol{h}}\|^4}\boldsymbol{H} + \frac{E[\chi_{ij}^2]}{\|\bar{\boldsymbol{h}}\|^2}\boldsymbol{I} \\
&\approx -\frac{2(\|\bar{\boldsymbol{h}}\|^2 \sigma_X^2 + \sigma_U^2 - \sigma_v^2)}{\|\bar{\boldsymbol{h}}\|^4 \sigma^2}\boldsymbol{I} + \frac{8(\sigma_U^2 - \sigma_v^2)\boldsymbol{H}}{\|\bar{\boldsymbol{h}}\|^6 \sigma^2}
\end{aligned} \tag{57}$$

$$\varpi = -\frac{2(\|\bar{\boldsymbol{h}}\|^2 \sigma_X^2 + \sigma_U^2 - \sigma_v^2)}{\|\bar{\boldsymbol{h}}\|^4 \sigma^2} \tag{58}$$

A combination of the above results with the stability conditions in Eq. (49) leads to an easily computable step-size constraint.

$$0 < \mu < \frac{\|\bar{\boldsymbol{h}}\|^4 \sigma^2}{\|\bar{\boldsymbol{h}}\|^2 \sigma_X^2 + \sigma_U^2 - \sigma_v^2} \tag{59}$$

## 5. Computational Complexity

In this section, the computational complexity of the proposed algorithms, along with several baseline methods, is analyzed in terms of Big-O notation. For consistency, let $L(4P+3)$ denote the number of input features after FLANN expansion, $N$ the total number of samples, and $M$ the window size used for quantization in QMTEE-FLANN.

1) LMS-FLANN: The LMS-FLANN algorithm updates the filter weights based on the instantaneous error. Each iteration requires: One inner product computation with a complexity of $W^T X$ with a complexity of $O(L(4P+3))$; A scalar error computation $O(1)$; A weight update operation $O(L(4P+3))$. Thus, the overall per-iteration complexity of LMS-FLANN is $O(L(4P+3))$, which is highly efficient and suitable for real-time applications.

2) MCC-FLANN: The MCC-FLANN algorithm introduces a nonlinear exponential weighting based on the error, with the update rule involving an additional expemtial function computation. Each iteration requires: One inner product $O(L(4P+3))$; An exponential computation $O(1)$; A weight update $O(L(4P+3))$. The per-iteration complexity remains: $O(L(4P+3))$, although the constant factor is higher due to the computational cost of the exponential function.

3) TLS-FLANN: In the TLS-FLANN framework, to properly account for noise in both input and output signals, a pseudoinverse or matrix decomposition is typically required. If performed batch-wise, the complexity of Singular Value Decomposition (SVD) for an $(L(4P+3)+1)\times N$ matrix is $O(\min\{(L(4P+3))^2 N, L(4P+3)N^2\})$. Assuming $N \gg L(4P+3)$, the complexity approximates to: $O((L(4P+3))^2 N)$. However, under an online or recursive update scheme, each iteration primarily involves small matrix inversions or updates, leading to a per-iteration complexity of approximately: $O((L(4P+3))^2)$.

4) MTEE-FLANN: The MTEE-FLANN algorithm integrates the advantages of MEE and TLS criteria under the EIV model. It involves estimating the error entropy, typically based on a Parzen window kernel density estimator. At each iteration, the error entropy calculation requires comparisons across all samples, resulting in a complexity of: $O(L(4P+3)M)$.

5) QMTEE-FLANN: In the QMTEE-FLANN algorithm, instead of computing the information potential over all pairwise error differences as in conventional MEE-based methods, only the relationship between the instantaneous error and the local mean of a sliding window is considered. Specifically, at each iteration, the following operations are performed: The sliding windows mean of recent $M$ errors, denoted as $\overline{e}_i$, is maintained. If optimized incremental updates are employed (e.g., using a cumulative sum and subtraction of the oldest sample), the computation of the sliding mean requires $O(1)$ operations. Otherwise, a full recomputation would incur $O(M)$ complexity. The instantaneous error $e_j$ is compared with $\overline{e}_i$, and the weight update involves a complexity of $O(L(4P+3))$ due to the dimensionality of the expanded input features.Thus, under optimized implementation, the overall per-iteration computational complexity of QMTEE-FLANN is $O(L(4P+3))$. If the sliding window mean is recomputed without optimization, the complexity becomes $O(L(4P+3)+M)$.

Compared to the original MTEE-FLANN algorithm with complexity $O(L(4P+3)M)$, the proposed QMTEE-FLANN significantly reduces the computational burden, making it highly suitable for real-time and large-scale applications.

Table 2 The Iterative Process of the MTEE-FLANN Algorithm.

| Algorithms | Complexity |
|---|---|
| LMS-FLANN | $O(L(4P+3))$ |
| MCC-FLANN | $O(L(4P+3))$ |
| TLS-FLANN | $O((L(4P+3))^2)$ |
| MTEE-FLANN | $O(L(4P+3)M)$ |
| QMTEE-FLANN | $O(L(4P+3)+M)$ |

## 6. Simulatiom

This section focuses on the application of the proposed algorithms in ANC. Part. 6.1 introduces the new nonlinear ANC under EIV model. In Part. 6.2, the performance comparison of FsFLMTEE and FsFLQMTEE under $S\alpha S$ noise is discussed, while Part. 6.3 and 6.4 focus on the simulation of realistic scenarios and explore the performance comparison of these algorithms under real-life server room noise and real-life factory floor noise, respectively.

*6.1. Nonlinear Active Noise Control under EIV Model*

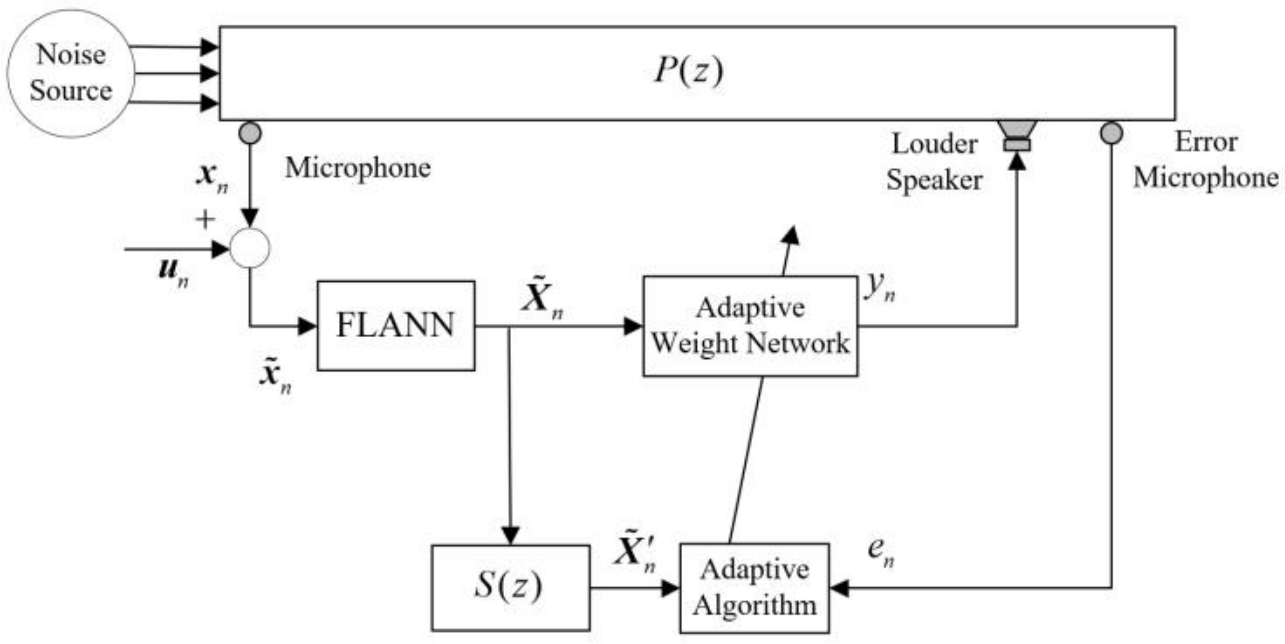

Figure. 3. .Nonlinear ANC system under the EIV model.

Fig. 3 demonstrates a nonlinear ANC system under the EIV model. The residual noise caused by the error microphone is below:

$$e_n = d_n - y_n * s_{N,n} \tag{60}$$

where the linear convolution operator $*$ is defined and the impulsive response of the secondary channel is denoted as $s_{N,n}$. The cost functions of the FcMTEE-FLANN and FcQMTEE-FLANN algorithms satisfy Eq. (61).

$$\boldsymbol{W}_{n+1} = \boldsymbol{W}_n + \mu \nabla J * s_{N,n} \tag{61}$$

In this experiment, the standard Averaged Noise Reduction (ANR) is employed as the evaluation metric, and its formulation is:

$$\mathrm{ANR}_n = 20\log\left(\frac{\mathrm{X}_{e,n}}{\mathrm{X}_{d,n}}\right) \tag{62}$$

$$\mathrm{X}_{e,n} = \varsigma \mathrm{X}_{e,n-1} + (1-\varsigma)\,|\,e_n\,|, \mathrm{X}_{e,0} = 0 \tag{63}$$

$$\mathrm{X}_{d,n} = \varsigma \mathrm{X}_{d,n-1} + (1-\varsigma)\,|\,y_n\,|, \mathrm{X}_{d,0} = 0 \tag{64}$$

where the smoothing parameter $\varsigma \to 1, \varsigma \neq 1$. Here $\varsigma$=0.999 .

*6.2. SαS Noise*

In this part, the reference noise is obtained by SαS noise [24], which is expressed as

$$\varphi_j = \exp\left\{ t\vartheta j - \varphi\,|\,j\,|^{\alpha}\,[1 + t\beta\,\mathrm{sign}(j)\omega(j,\alpha)]\right\} \tag{65}$$

where

$$\omega(j,\alpha) = \begin{cases} \frac{\pi}{2}\log|\,j\,|, & \alpha = 1 \\ \tan\frac{\alpha\pi}{2}, & \alpha \neq 1 \end{cases} \tag{66}$$

$$\mathrm{sign}(j) = \begin{cases} 1, & j > 0 \\ 0, & j = 0 \\ -1, & j < 0 \end{cases} \tag{67}$$

The SαS noise's parameters are $\beta$=0,$\varphi$=1,$\vartheta$=0 , and the standard distribution's characteristic function is

$$\varphi_j = \exp\left(-\,|\,j\,|^{\alpha}\right) \tag{68}$$

The intensity of the noise is controlled by adjusting the $\alpha$ -parameter. Smaller values of $\alpha$ correspond to higher noise intensity Gaussian noise with zero mean and a variance of 0.01 is chosen for both input and output noise, i.e., $\boldsymbol{u}_n$ and $v_n$. The nonlinear forward path is:

$$d_n = u_{n-2} + \delta_1 u_{n-2}^2 - \delta_2 u_{n-1}^3 \tag{69}$$

The functions of $\delta_i$, i=1,2,... are used to assess the intensity of Eq. (69). The experiment is configured with $\delta_1$=0.08 and $\delta_2$=0.04, where $\mathbf{x}_n$ represents the $\alpha$-stable distributed input described earlier. In Eq. (69), $u_n = \mathbf{x}_n * \rho_n$, where $\rho_n$ represents the impulse response of the primary channel transfer function. The transfer function of the primary channel Eq. (70) and the secondary channel with minimum phase Eq. (71) and non-minimum phase Eq. (72) characteristic are:

$$\mathrm{P}_z = z^{-3} - 0.3z^{-4} + 0.2z^{-5} \tag{70}$$

$$S_z = z^{-2} + 0.5z^{-3} \tag{71}$$

$$S_z = z^{-2} + 1.5z^{-3} - z^{-4} \tag{72}$$

Fig.4 and 5 compare the performance of the FsFLLMS, FsFLTLS, FsFLMCC, FcFLMTEE, and FsFLQMTEE algorithms under varying strengths of SαS noise (1.9, 1.8, 1.7, and 1.6). Parameter selection is shown in Table 3. As shown in Fig. 4, algorithms that do not incorporate the TLS criterion become progressively less robust as the noise intensity increases and the FsFLLMS algorithm starts to diverge when the noise intensity reaches 1.6. Compare Fig.4 and 5, under the influence of alpha-stable (non-Gaussian) noise, MEE-based algorithms (FsFLMTEE and FsFLQMTEE) demonstrate significantly superior robustness. Additionally, the FsFLQMTEE algorithm outperforms the FsFLMTEE algorithm when a bias is introduced. A comparison between FsFLMTEE and FsFLQMTEE shows that the proposed new algorithms exhibit similar performance in the non-minimum phase secondary path. Overall, under the minimum phase and non-minimum phase, the algorithms maintain excellent robustness as noise intensity increases.

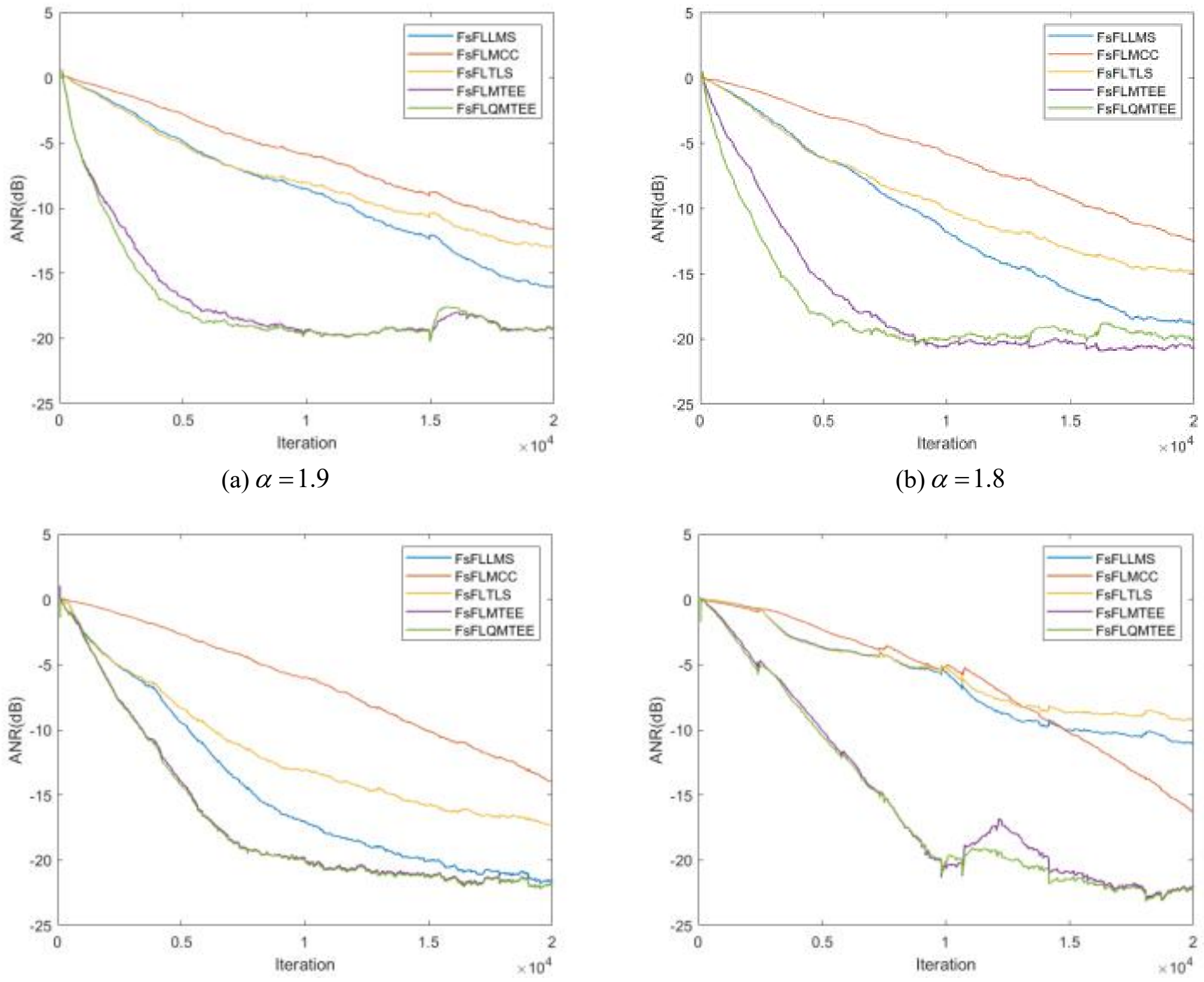

(a) $\alpha = 1.9$ (b) $\alpha = 1.8$

(c) $\alpha = 1.7$ (d) $\alpha = 1.6$

Figure. 4. The ANR curves with minimum phase secondary path under SαS interference.

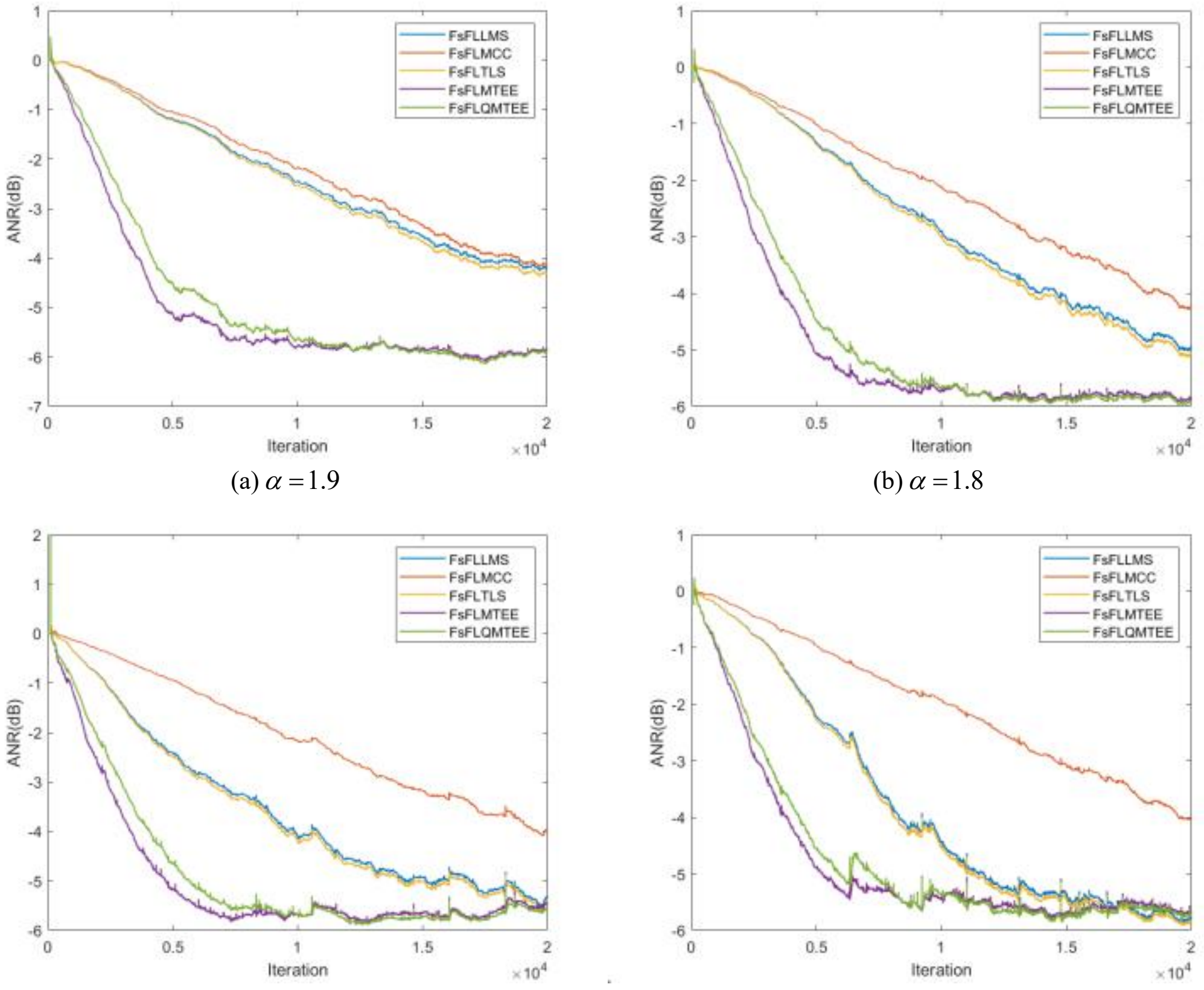

(a) $\alpha=1.9$ (b) $\alpha=1.8$

(c) $\alpha=1.7$ (d) $\alpha=1.6$

Figure. 5. The ANR curves with non-minimum phase secondary path under SαS interference.

Table 3 Choice of parameters

| Algorithms | | FsFLTLS | FsFLTLS | FsFLMCC | FsFLMTEE | FsFLQMTEE |
|---|---|---|---|---|---|---|
| $\alpha$=1.9 | Min | $\mu=1\times10^{-4}$ | $\mu=1\times10^{-4}$ | $\mu=1\times10^{-4},\sigma_1=2$ | $\mu=0.03,\sigma_2=10$ | $\mu=0.05,\sigma_2=10$ |
| | Non-Min | $\mu=5\times10^{-6}$ | $\mu=5\times10^{-6}$ | $\mu=1\times10^{-5},\sigma_1=2$ | $\mu=0.001,\sigma_2=7$ | $\mu=0.0015,\sigma_2=7$ |
| $\alpha$=1.8 | Min | $\mu=1\times10^{-4}$ | $\mu=1\times10^{-4}$ | $\mu=1\times10^{-4},\sigma_1=2$ | $\mu=0.01,\sigma_2=5$ | $\mu=0.04,\sigma_2=5$ |
| | Non-Min | $\mu=5\times10^{-6}$ | $\mu=5\times10^{-6}$ | $\mu=1\times10^{-5},\sigma_1=2$ | $\mu=0.001,\sigma_2=7$ | $\mu=0.0015,\sigma_2=7$ |
| $\alpha$=1.7 | Min | $\mu=1\times10^{-4}$ | $\mu=1\times10^{-4}$ | $\mu=1\times10^{-4},\sigma_1=2$ | $\mu=0.01,\sigma_2=7$ | $\mu=0.02,\sigma_2=7$ |
| | Non-Min | $\mu=5\times10^{-6}$ | $\mu=5\times10^{-6}$ | $\mu=1\times10^{-5},\sigma_1=2$ | $\mu=0.001,\sigma_2=7$ | $\mu=0.0015,\sigma_2=7$ |
| $\alpha$=1.6 | Min | $\mu=1\times10^{-4}$ | $\mu=1\times10^{-4}$ | $\mu=1\times10^{-4},\sigma_1=2$ | $\mu=0.005,\sigma_2=7$ | $\mu=0.005,\sigma_2=7$ |
| | Non-Min | $\mu=5\times10^{-6}$ | $\mu=5\times10^{-6}$ | $\mu=1\times10^{-5},\sigma_1=2$ | $\mu=0.0005,\sigma_2=5$ | $\mu=0.0008,\sigma_2=5$ |

### *6.3. Real-life server room Noise*

This section evaluates the performance of the presented algorithms in a real-world noise environment. To evaluate their effectiveness, real server room (SR) noise is used as the reference noise. The signal segment comprises 20,000 samples with a sampling rate of 16 kHz. SR noise cancellation is a significant area of research in ANC [25]. In this case, we use an SR recording as the reference noise. The nonlinear forward path follows the model in Eq. (69) with $\rho_n$ derived using a 32-length FIR filter [26]. The minimum-phase secondary path is identical to Eq. (71). The rest of the simulation setup remains consistent with Part. 6.2.

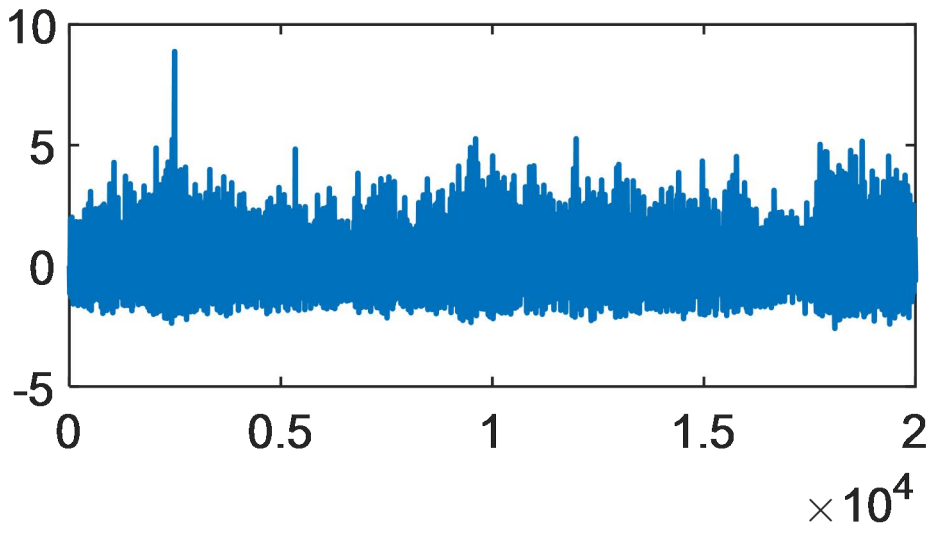

Figure. 6. The real-life server room noise.

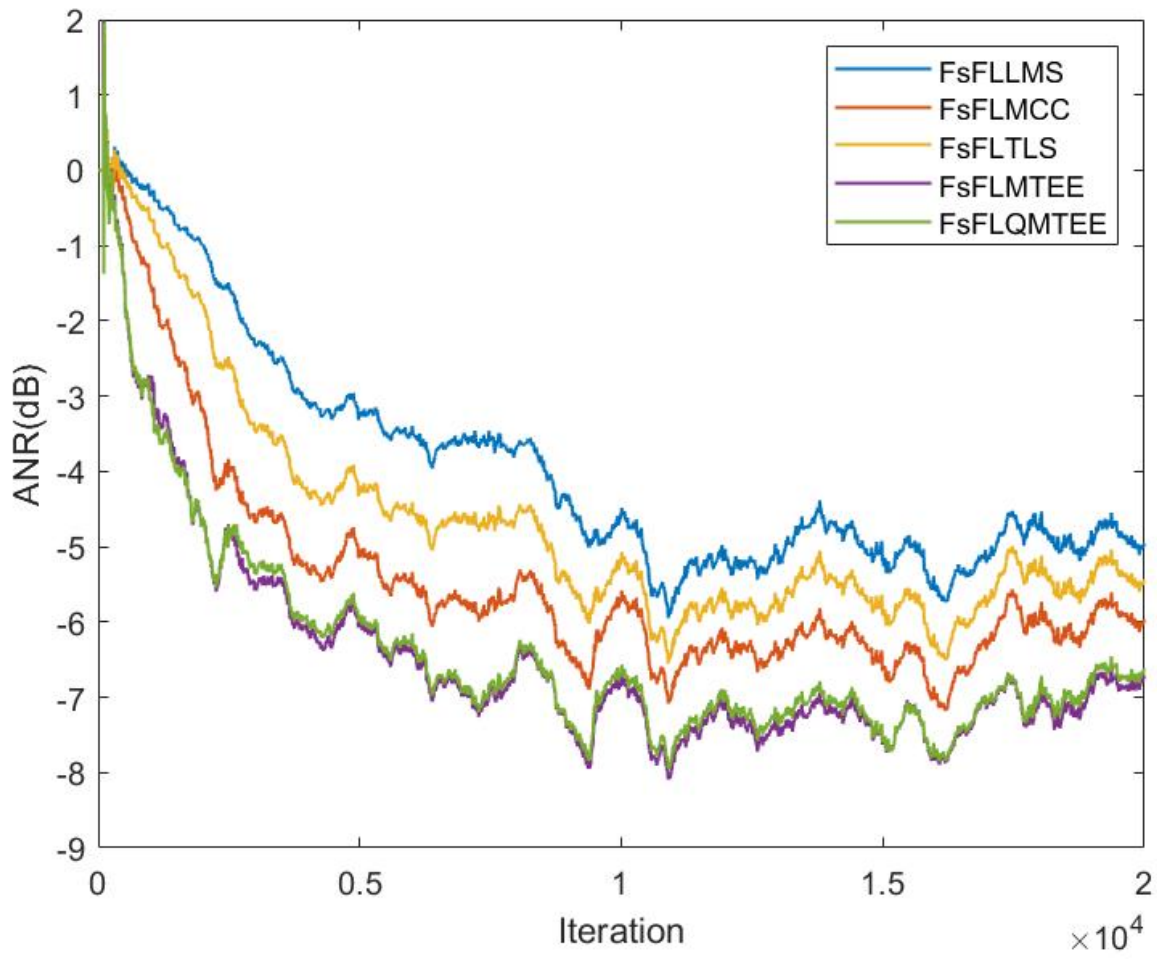

Figure. 7. The ANR curves under SR noise interference.

Fig. 6 shows the real-life SR noise. In Fig.7, these new algorithms on the basis of the FLANN filter model achieve lower steady-state errors compared to others. Among these algorithms, these proposed algorithms performs the best and the FsFLQMTEE shows the fastest convergence rate. Parameter selection is shown in Table 4.

Table 4 Choice of parameters

| Algorithms | Real-Life SR Noise Control | Real-Life FF Noise Control |
|---|---|---|
| FsFLLMS | $\mu$=1×10$^{-5}$ | $\mu$=1×10$^{-6}$ |
| FsFLMCC | $\mu$=5×10$^{-5}$,$\sigma_1$=3 | $\mu$=1×10$^{-5}$,$\sigma_1$=3 |
| FsFLTLS | $\mu$=2×10$^{-5}$ | $\mu$=2×10$^{-6}$ |
| FsFLMTEE | $\mu$=3×10$^{-3}$,$\sigma_2$=5 | $\mu$=4×10$^{-4}$,$\sigma_2$=12 |
| FsFLQMTEE | $\mu$=5×10$^{-3}$,$\sigma_2$=5 | $\mu$=6×10$^{-4}$,$\sigma_2$=12 |

*6.4. Real-life server room Noise*

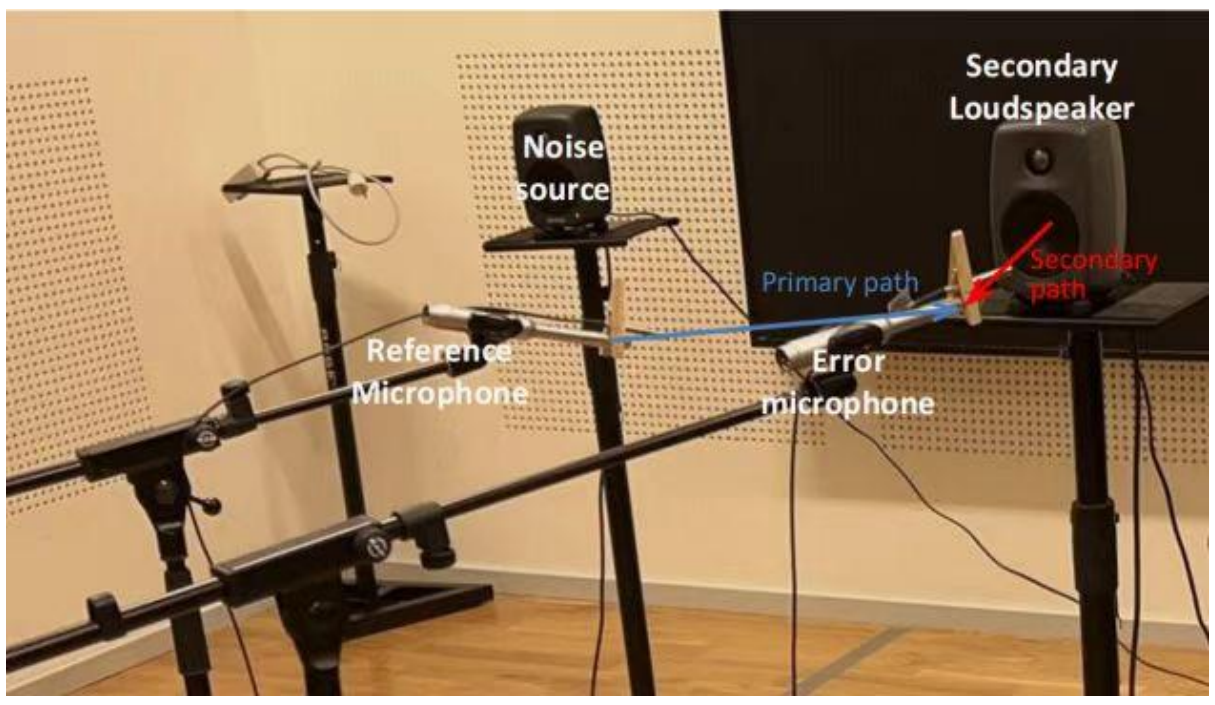

Figure. 8. Experimental environment setup.

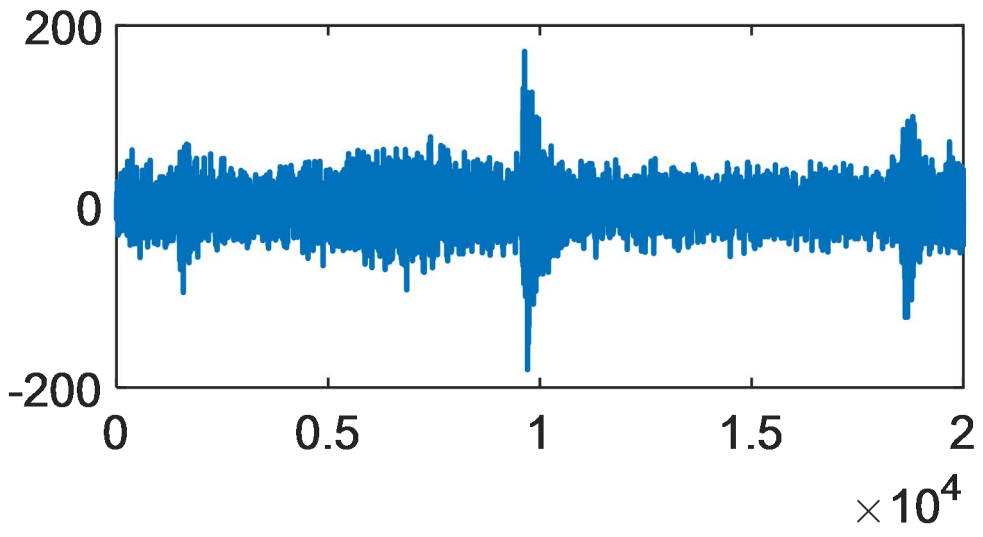

Figure. 9. The real-life factory floor noise.

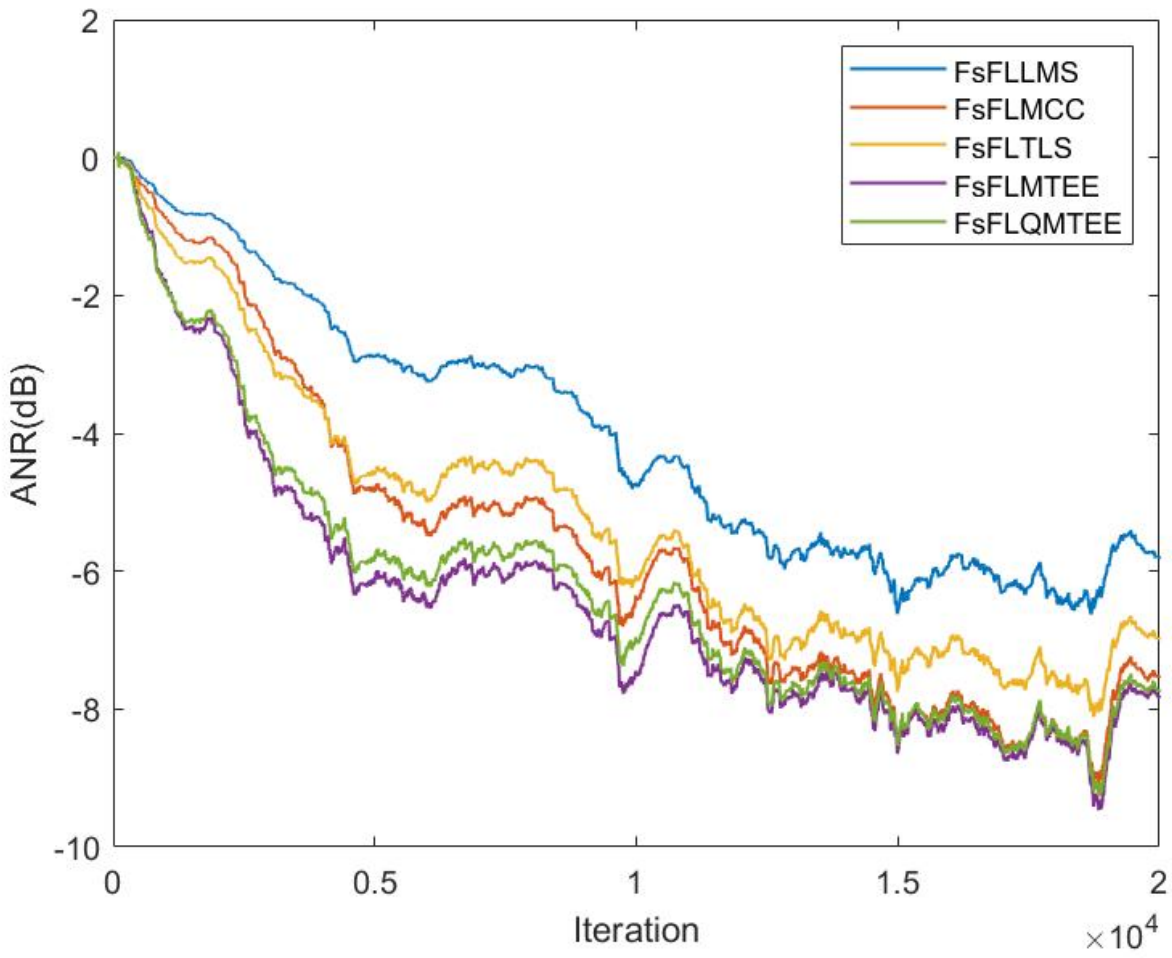

Figure. 10. The ANR curves under FF noise interference.

In this section, the reference noise, Factory Floor (FF) noise, is used to assess the accomplishment of these presented methods. The noise was recorded using a Bruel & Kjaer condenser microphone, with the samples subsequently captured onto digital audio tape. This 10-second recording was made near plate-cutting and electrical welding machines. During the recording process, the audio was downsampled to an 8 kHz sampling rate, with a 16-bit analog-to-digital converter used for quantization, and an anti-aliasing filter was applied. The impulsive response of the acoustic paths was assessed in a 4x5x6 meter acoustic lab at the Audio Analysis Lab, Aalborg University [8]. Fig. 9 shows the experimental setup, where two speakers were used—one as the noise source and the other as the secondary source. The reference microphone was positioned 1.5 meters away from the noise source, with the microphone height set at 1.2 meters.

In contrast to the model in [8], this study considers a scenario where the microphone picks up noise while receiving the signal. Notably, no pre-emphasis stage was used in this setup. The time-domain waveform of the FF noise is displayed in Fig. 9. In Fig. 10, the FsFLMCC diverges when the noise power increases significantly and takes a long time to converge. On the other hand, the FsFLLMS, FsFLTLS, FsFLMTEE, and FsFLQMTEE algorithms show better robustness and tracking performance, with the proposed algorithms consistently providing the best robustness.

## 7. Conclusion

This work pioneers the integration of nonlinear Errors-In-Variables (EIV) models into Active Noise Control (ANC) system modeling, addressing a critical limitation of conventional linear assumptions in realistic environments. To effectively process the resulting nonlinear and uncertain system dynamics, a novel MTEE-FLANN algorithm was proposed, alongside its computationally optimized variant, QMTEE-FLANN. A comprehensive theoretical analysis was conducted, covering extremum point behavior, stabilization conditions, and computational complexity. Crucially, by reformulating the ANC problem within the nonlinear EIV framework, this paper establishes a fundamentally new modeling approach for adaptive noise control. Extensive simulations validate the proposed framework, demonstrating that the FsFLMTEE and FsFLQMTEE algorithms achieve superior noise reduction performance under non-Gaussian noise conditions across various real-world scenarios. These results confirm the effectiveness of the proposed EIV-based ANC modeling and algorithmic framework, offering a new thinking for the design of robust adaptive filtering systems in practical noisy environments.

## 8. Acknowledgement

This work was partially supported by National Natural Science Foundation of China [grant numbers 62171388, 61871461, 61571374, 62201097].